\documentclass[aps,prc,twocolumn,superscriptaddress,showpacs,preprintnumbers,nofootinbib]{revtex4-1}
\usepackage{graphicx,amsmath,amssymb,bm,xcolor,tabularx}
\usepackage{enumerate}
\usepackage{multirow,dcolumn}
\usepackage{color}

\newcommand{\fm}{\,\mathrm{fm}}

\newcommand{\beq}{\begin{equation}}
\newcommand{\eeq}{\end{equation}}
\newcommand{\beqa}{\begin{eqnarray}}
\newcommand{\eeqa}{\end{eqnarray}}

\begin{document}

\preprint{INT-PUB-18-014}

\title{A critical examination of constraints on the equation of state of dense matter obtained from GW170817}

\author{I.\ Tews}
\email[E-mail:~]{itews@uw.edu}
\affiliation{Institute for Nuclear Theory,
University of Washington, Seattle, WA 98195-1550, USA}
\affiliation{JINA-CEE, Michigan State University, 
East Lansing, MI, 48823, USA}
\author{J.\ Margueron}
\email[E-mail:~]{jmargue@uw.edu}
\affiliation{Institute for Nuclear Theory,
University of Washington, Seattle, WA 98195-1550, USA}
\affiliation{Institut de Physique Nucl\'eaire de Lyon, CNRS/IN2P3, Universit\'e de Lyon, Universit\'e Claude Bernard Lyon 1, F-69622 Villeurbanne Cedex, France}
\author{S.\ Reddy}
\email[E-mail:~]{sareddy@uw.edu}
\affiliation{Institute for Nuclear Theory,
University of Washington, Seattle, WA 98195-1550, USA}
\affiliation{JINA-CEE, Michigan State University, 
East Lansing, MI, 48823, USA}
 
\begin{abstract}
The correlation of the tidal polarizabilities $\Lambda_1$-$\Lambda_2$ for GW170817 is predicted by combining dense-matter equations of state (EOSs) that satisfy nuclear physics constraints with the chirp mass and mass asymmetry for this event.
Our models are constrained by calculations of the neutron-matter EOS using chiral effective field theory Hamiltonians with reliable error estimates  up to once or twice the nuclear saturation density. In the latter case, we find that GW170817 does not improve our understanding of the EOS.
We contrast two distinct extrapolations to higher density: a minimal model (MM) which assumes that the EOS is a smooth function of density described by a Taylor expansion and a more general model parametrized by the speed of sound that admits phase transitions. This allows us to identify regions in the $\Lambda_1$-$\Lambda_2$ plots that could favor the existence of new phases of matter in neutron stars. We predict the combined tidal polarizability of the two neutron stars in GW170817 to be $80\le \tilde{\Lambda}\le 580$ ($280\le \tilde{\Lambda}\le 480$ for the MM), which is smaller than the range suggested by the LIGO-Virgo data analysis. 
Our analysis also shows that GW170817 requires a NS with $M=1.4M_\odot$ to have a radius $9.0<R_{1.4}<13.6$~km ($ 11.3  <R_{1.4}< 13.6$~km for the MM). 
\end{abstract}

 \maketitle
 
 \section{Introduction}
 
The first multimessenger observations of a binary neutron-star (NS) merger, GW170817, marks the beginning of a new era in astronomy~\cite{Abbott2017b}. On August 17, 2017, gravitational waves (GW) from this event were observed in the interferometer network consisting of Advanced LIGO and Virgo~\cite{Abbott:2017}. 1.7 s after the GW signal, detectors onboard Fermi and Integral observed a short gamma-ray burst, and after hours to days optical and infrared observations by several ground-based telescopes revealed emissions that were consistent with the decay of heavy nuclei produced and ejected during the merger~\cite{Abbott2017b}. This event and more expected in future have the potential to unravel many unsolved questions related to NSs, their interior composition, and their role in nucleosynthesis and astrophysics.  

In this article, we confront the results of the LIGO-Virgo (LV) data analysis~\cite{Abbott:2018wiz} (updating the analysis presented in Ref.~\cite{Abbott:2017}), with our own analysis using equations of state (EOSs) consistent with the current understanding of nuclear interactions and the properties of nucleonic matter in the vicinity of nuclear saturation density, $n_{\text{sat}}~=~0.16 \fm^{-3}$.  Unlike the LV analysis, we assume that both compact objects are governed by the same EOS, excluding, e.g., a binary black-hole NS scenario. We employ two models for the high-density EOS: a minimal model (MM) which relies on an extrapolation of a parametrized nuclear EOS guided by a density expansion about $n_{\text{sat}}$, and a second model, which we call speed-of-sound model (CSM), which is a general parametrization of the speed of sound $c_S$ and includes phase transitions. Both models are constrained by calculations of pure neutron matter (PNM) up to a transition density $n_{\text{tr}}$, which is chosen to be either $n_{\text{sat}}$ or $2~n_{\text{sat}}$. The neutron-matter EOS, obtained by solving the many-body Hamiltonian derived from chiral effective field theory (EFT) using Quantum Monte Carlo (QMC) methods, is expected to provide reliable error estimates up to $2n_{\text{sat}}$~\cite{Tews:2018kmu}. Our models are also constrained to be stable (pressure and $c_S$ are positive), causal ($c_S<c$, with the speed of light $c$), and are required to support a NS maximum mass $M_{\text{max}} \ge 1.9~M_\odot$. Our analysis, which contrasts these models in light of the data from GW170817, allows us to: (i) establish the range of tidal polarizabilities $\Lambda$ that are predicted by an EFT-based nuclear EOS expected to describe matter up to $2n_{\text{sat}}$ with associated errors, (ii) identify correlations between $\Lambda_1$-$\Lambda_2$ that are sensitive to the properties of matter between $n_{\text{sat}}$ and $2n_{\text{sat}}$, and to those at higher densities, where reliable calculations do not currently exist; and (iii) contrast extrapolations of nuclear EOSs to those in which more extreme variations, including phase transitions, may be present and thereby address the masquerade problem~\cite{Alford:2004pf}. 

\begin{figure*}[t]
\includegraphics[trim= 0.1cm 0.1cm 0.8cm 0.8cm, clip=,width=0.67\columnwidth]{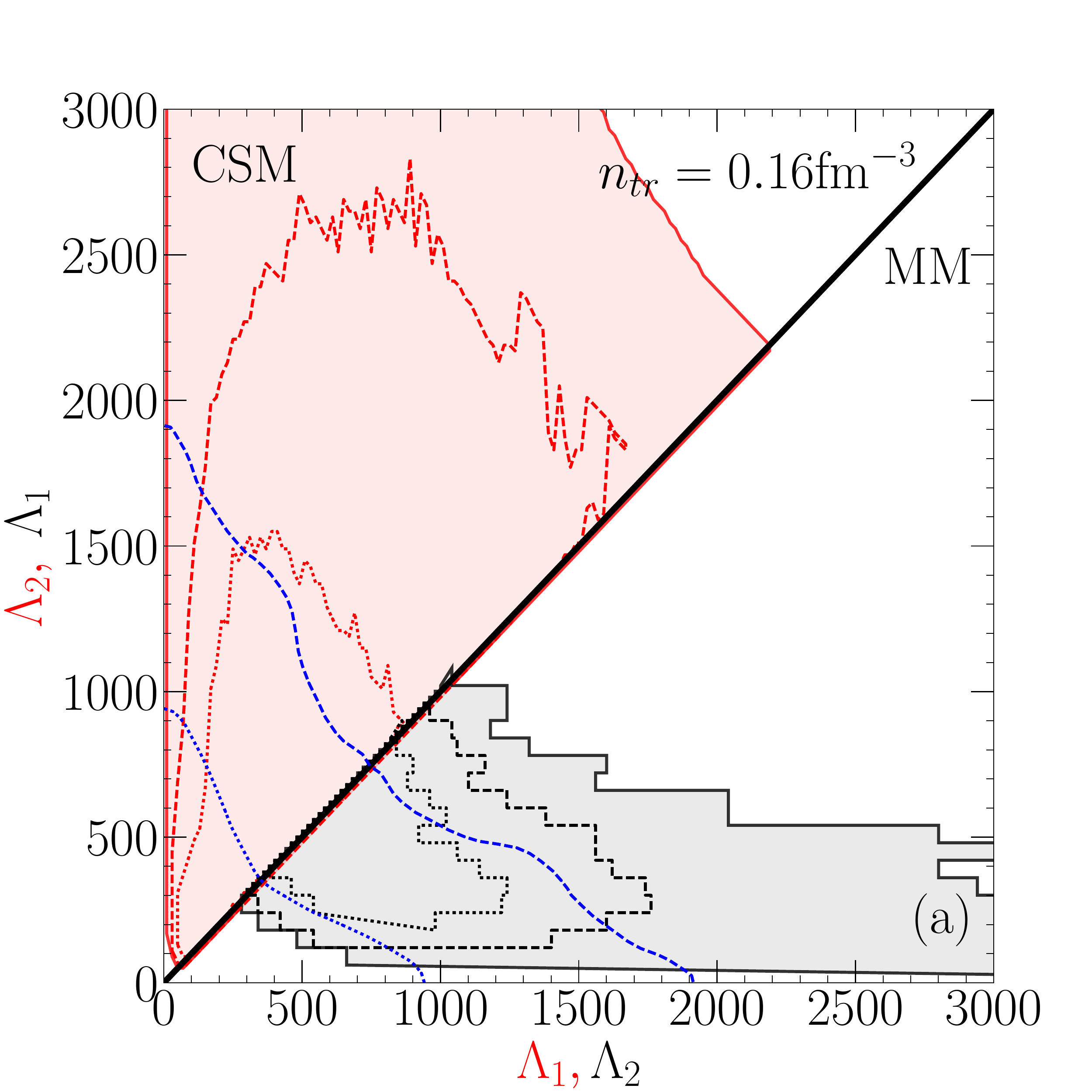}
\includegraphics[trim= 0.1cm 0.1cm 0.8cm 0.8cm, clip=,width=0.67\columnwidth]{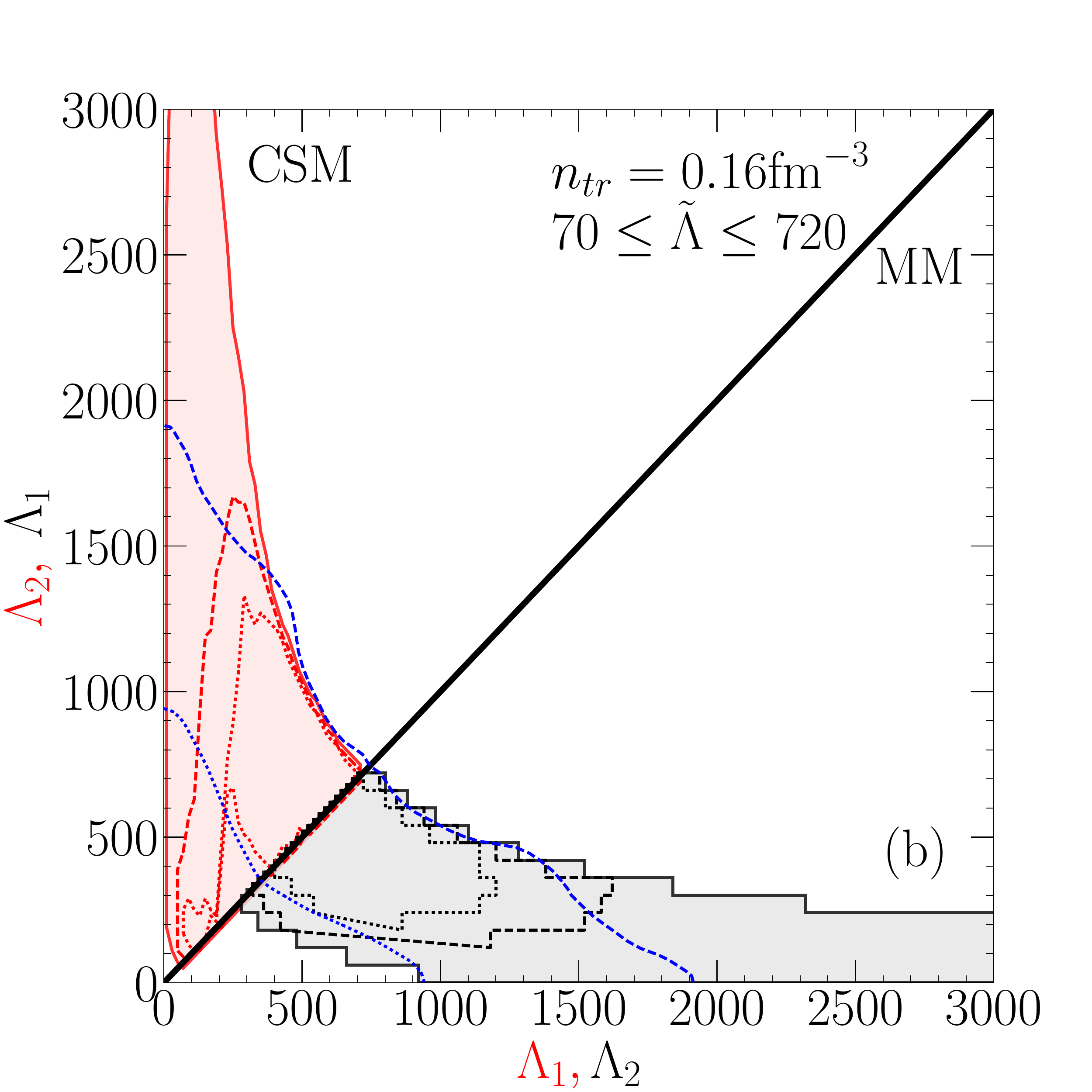}
\includegraphics[trim= 0.1cm 0.1cm 0.8cm 0.8cm, clip=,width=0.67\columnwidth]{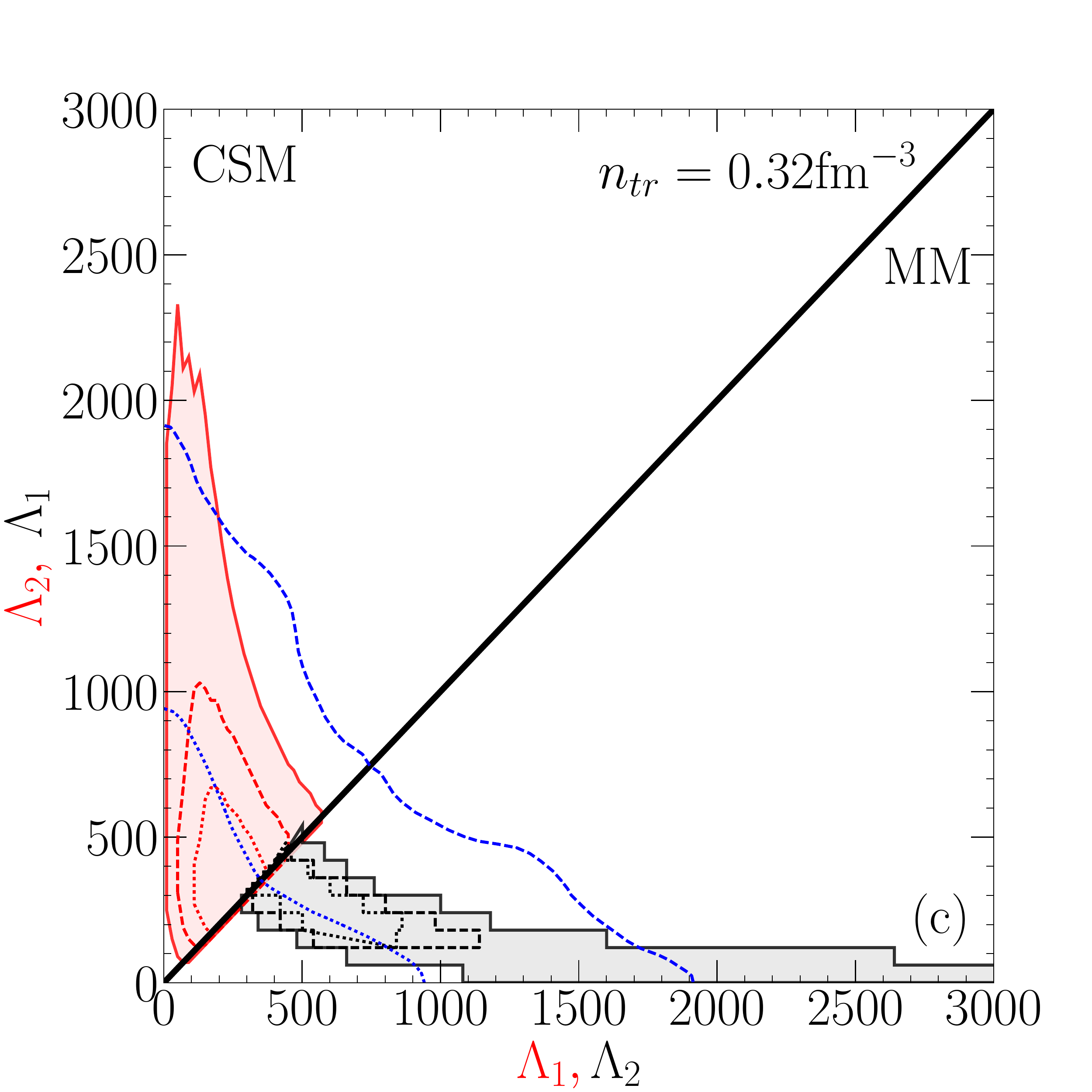}
\caption{\label{fig:LamLam}
Envelopes for the CSM (red) and the MM (black) for the correlation of $\Lambda_1$ and $\Lambda_2$ for the two NSs in GW170817. We show results (a) for $n_{\text{tr}}=n_{\text{sat}}$ and no constraint on $\tilde{\Lambda}$, (b) for $n_{\text{tr}}=n_{\text{sat}}$ when additionally enforcing $70\le \tilde{\Lambda}\le 720$, and (c) for $n_{\text{tr}}=2 n_{\text{sat}}$ and no constraint on $\tilde{\Lambda}$. We also show $90\%$ (dashed lines) and $50\%$ (dotted lines) probability contours for the MM and the CSM, and compare with the corresponding $90\%$ and $50\%$ contours from the LV analysis (blue lines).}
\end{figure*} 

We present our main findings in Fig.~\ref{fig:LamLam}, where we show correlation plots of the tidal polarizabilities $\Lambda_1$ and $\Lambda_2$ of the two neutron stars in GW170817. We find that our analysis including pure neutron-matter constraints up to $n_\text{sat}$ and the assumption that the neutron stars are governed by the same EOS is compatible with the LV analysis. The CSM, however, allows for a wider range of $\Lambda_i$ [see Fig.~\ref{fig:LamLam}(a)]. In this case, the LV analysis provides useful constraints on the EOS at supra-nuclear densities [see Fig.~\ref{fig:LamLam}(b) where $70\le \tilde{\Lambda}\le 720$ is imposed].  However, if the neutron-matter EOS is assumed to be valid up to $2n_\text{sat}$, the predicted $\Lambda_i$ are more tightly constrained than suggested by the LV analysis, as is evident from Fig.~\ref{fig:LamLam}(c). In this case, GW170817 does not provide new insights about the EOS and future observations would need to achieve an uncertainty $\Delta \tilde{\Lambda}\approx 300-400$ to do so. Furthermore, if the observational uncertainty reaches a precision of $\Delta \tilde{\Lambda} < 200$, a comparison between CSM and MM predictions in Fig.~\ref{fig:LamLam}(c) suggests that it would be possible to probe phase transitions at supra-nuclear density. 

This paper is structured as follows: In Sec.~\ref{sec:EOS} we introduce the EOS models we use in this work. In Sec.~\ref{sec:GW170817} we study the implications of the recent observation of GW170817. Finally, we summarize in Sec.~\ref{sec:Summary}.

\section{Equation-of-state models}\label{sec:EOS}
\subsection{Low-density neutron-matter constraints}

Low-density neutron-matter is rather well determined by microscopic approaches based on Hamiltonians constrained by $NN$ scattering data and information on light nuclei. At very low densities, PNM is close to the unitary limit where the large two-body scattering length dominates. Here, measurements of cold atomic gases~\cite{Nascimbene2010,Zwierlein:2015} validate the QMC methods used in this work~\cite{Carlson:2008zza}. In addition, two- and many-body interactions in neutron matter are simpler than in systems containing also protons~\cite{Hebeler:2009iv}. Therefore, current many-body methods are well suited to describe neutron matter, and predictions of the neutron-matter EOS based on different many-body approaches and Hamiltonians are typically in very good agreement~\cite{Gandolfi:2015jma, Hebeler:2015hla}. 

Among them, chiral EFT~\cite{Weinberg1979, Weinberg1990, Weinberg1991, Epelbaum2009, Machleidt:2011zz} is a systematic and successful theory of nuclear forces that provides a well-defined prescription to estimate uncertainties. In this work, we consider neutron-matter results of Ref.~\cite{Tews:2018kmu}, which were obtained by using chiral EFT Hamiltonians from Refs.~\cite{Gezerlis:2013ipa,Lynn:2015jua}. These Hamiltonians have been tested in light- to medium-mass nuclei and in $n$-$\alpha$ scattering with great success~\cite{Lynn:2015jua, Lonardoni:2017hgs}. They also agree with experimental knowledge of the empirical parameters for the symmetry energy~\cite{Kolomeitsev:2016sjl, Margueron:2017eqc}. In Ref.~\cite{Tews:2018kmu}, the order-by-order convergence of these Hamiltonians was investigated, and it was found that the convergence was reasonable and consistent with power-counting arguments up to $2n_{\text{sat}}$ within theoretical uncertainty estimates. Thus, we base all our models on this neutron-matter EOS up to $n_{\text{tr}}$, which we vary to be $n_{\text{sat}}$ or $2n_{\text{sat}}$. 

\begin{figure*}[t]
\includegraphics[trim= 0.0cm 0 0 0, clip=,width=0.65\columnwidth]{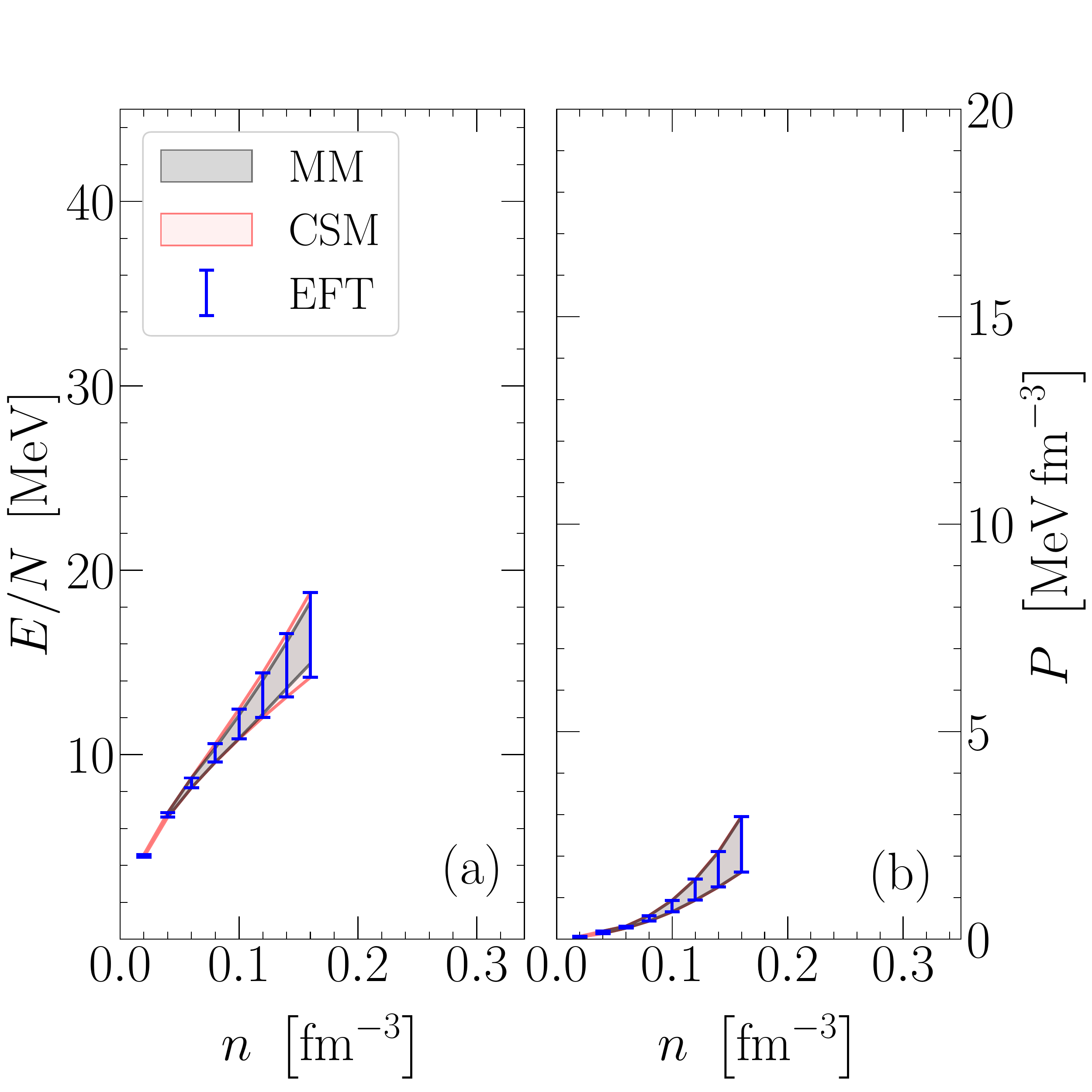}
\includegraphics[trim= 0.0cm 0 0 0, clip=,width=0.65\columnwidth]{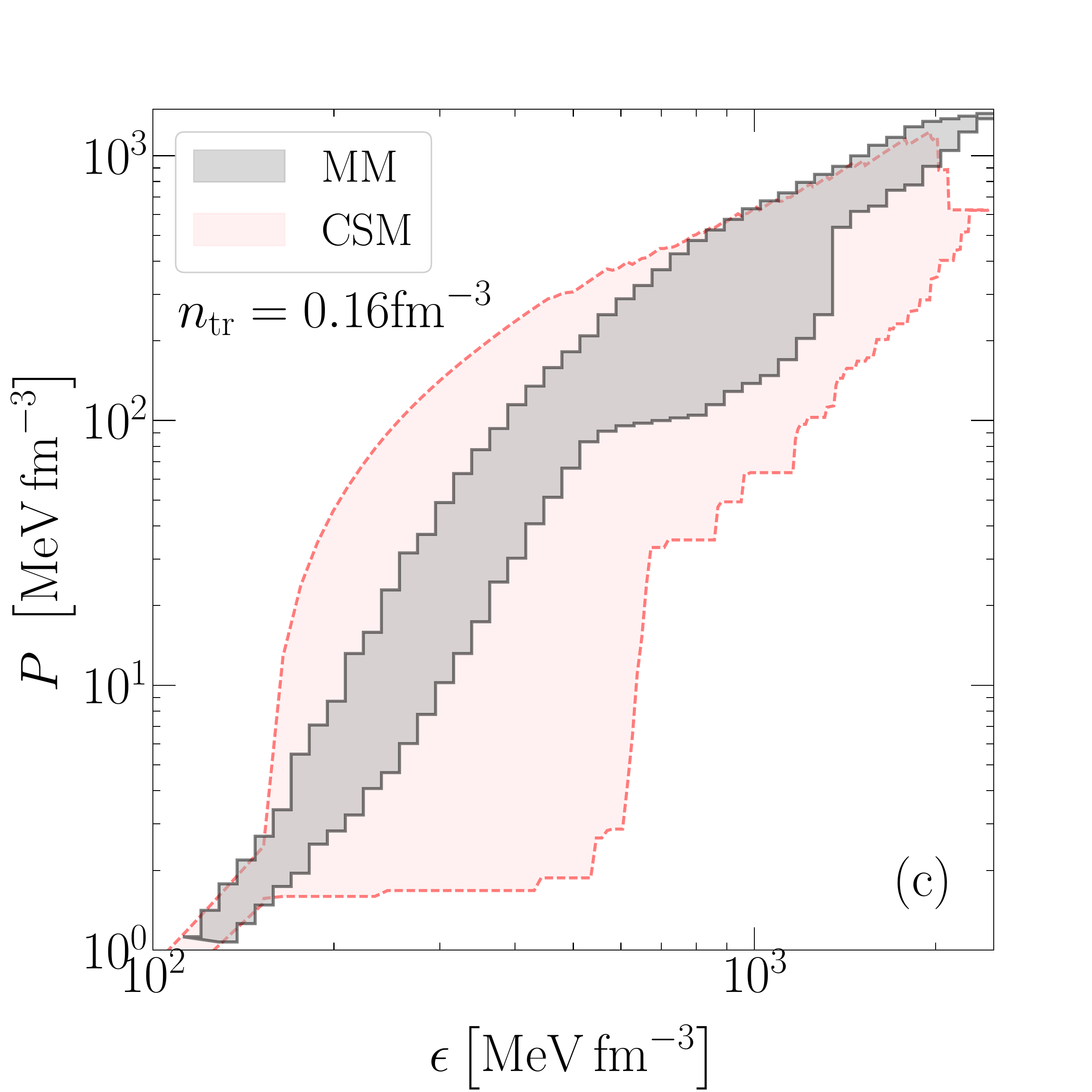}
\includegraphics[trim= 0.0cm 0 0 0, clip=,width=0.65\columnwidth]{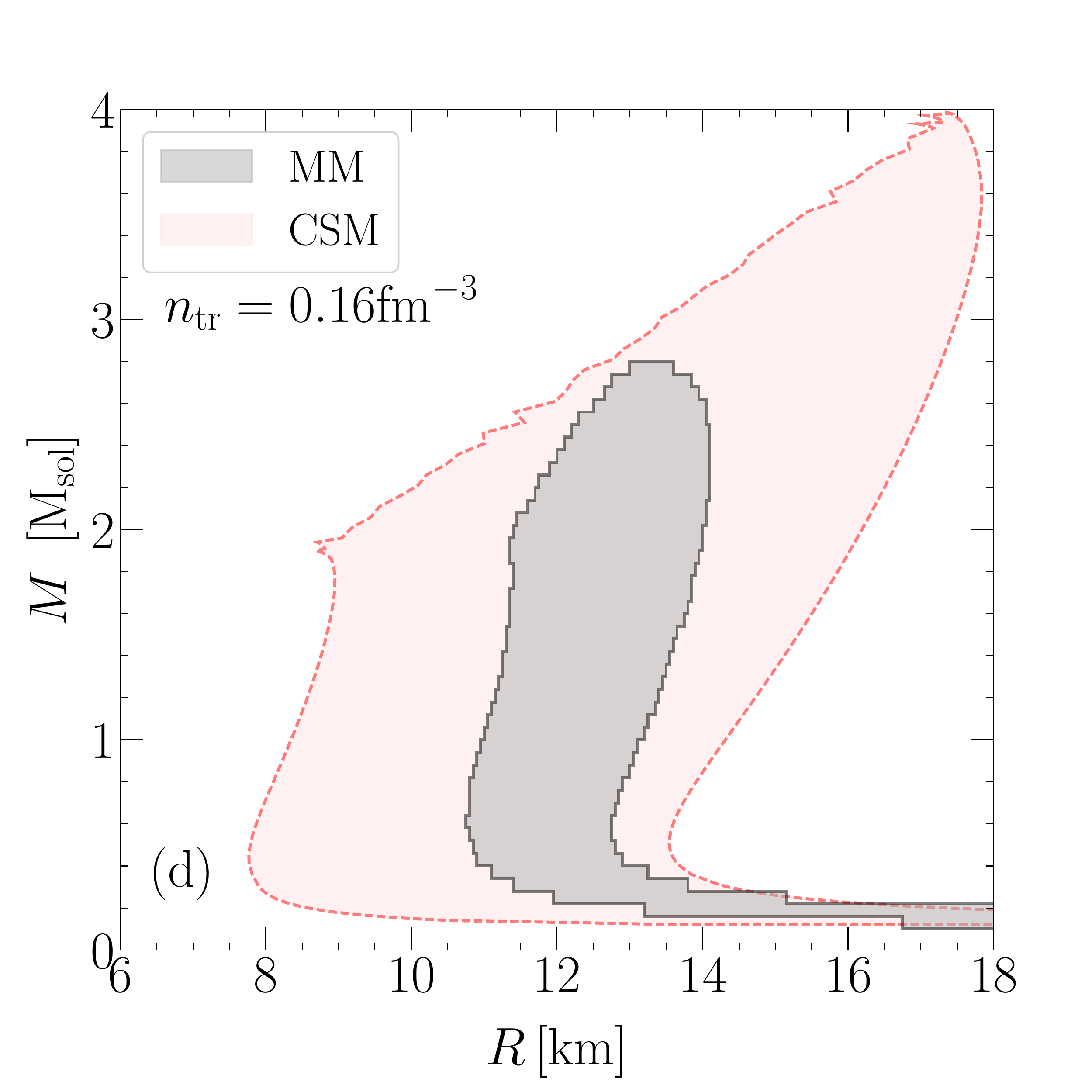}\\
\includegraphics[trim= 0.0cm 0 0 0, clip=,width=0.65\columnwidth]{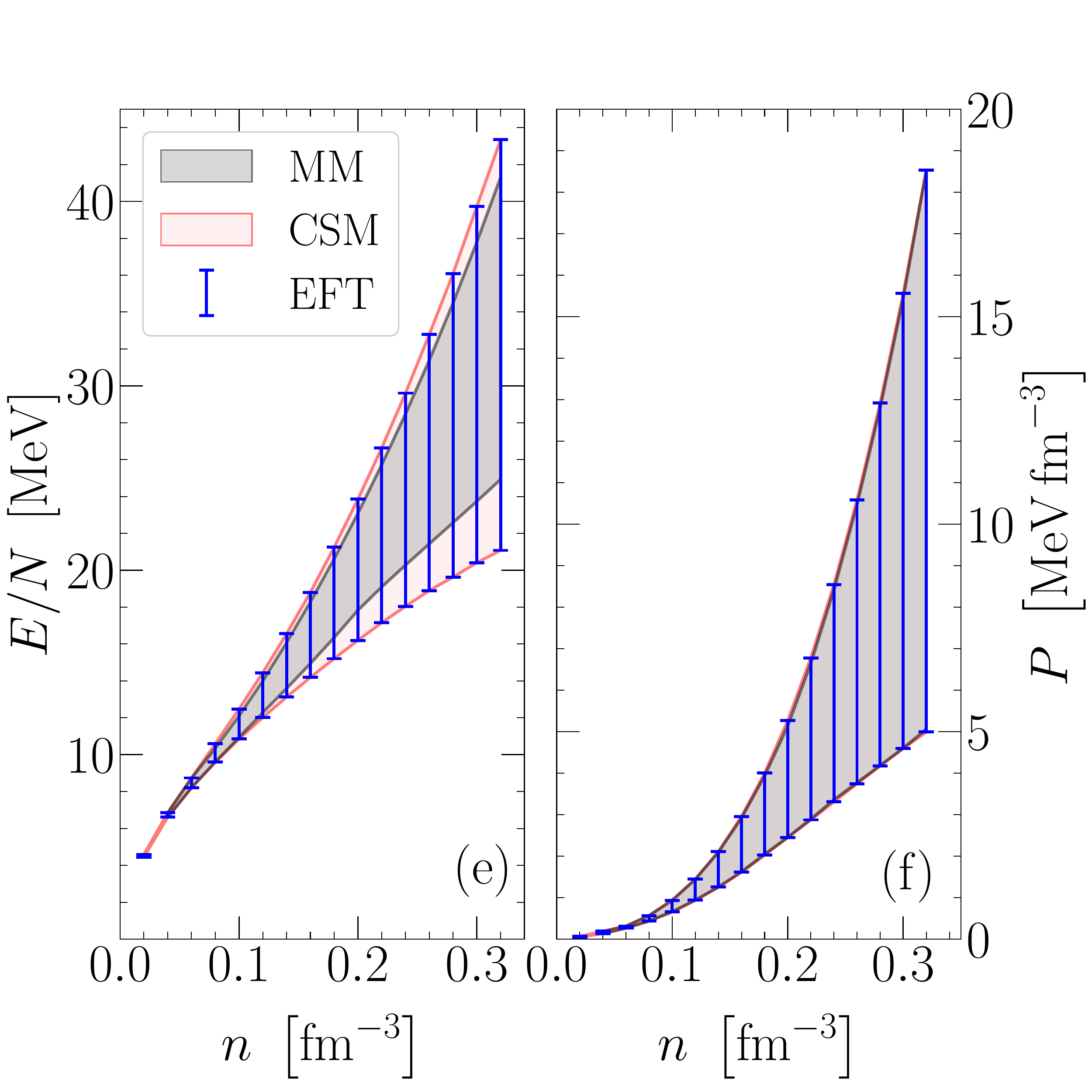}
\includegraphics[trim= 0.0cm 0 0 0, clip=,width=0.65\columnwidth]{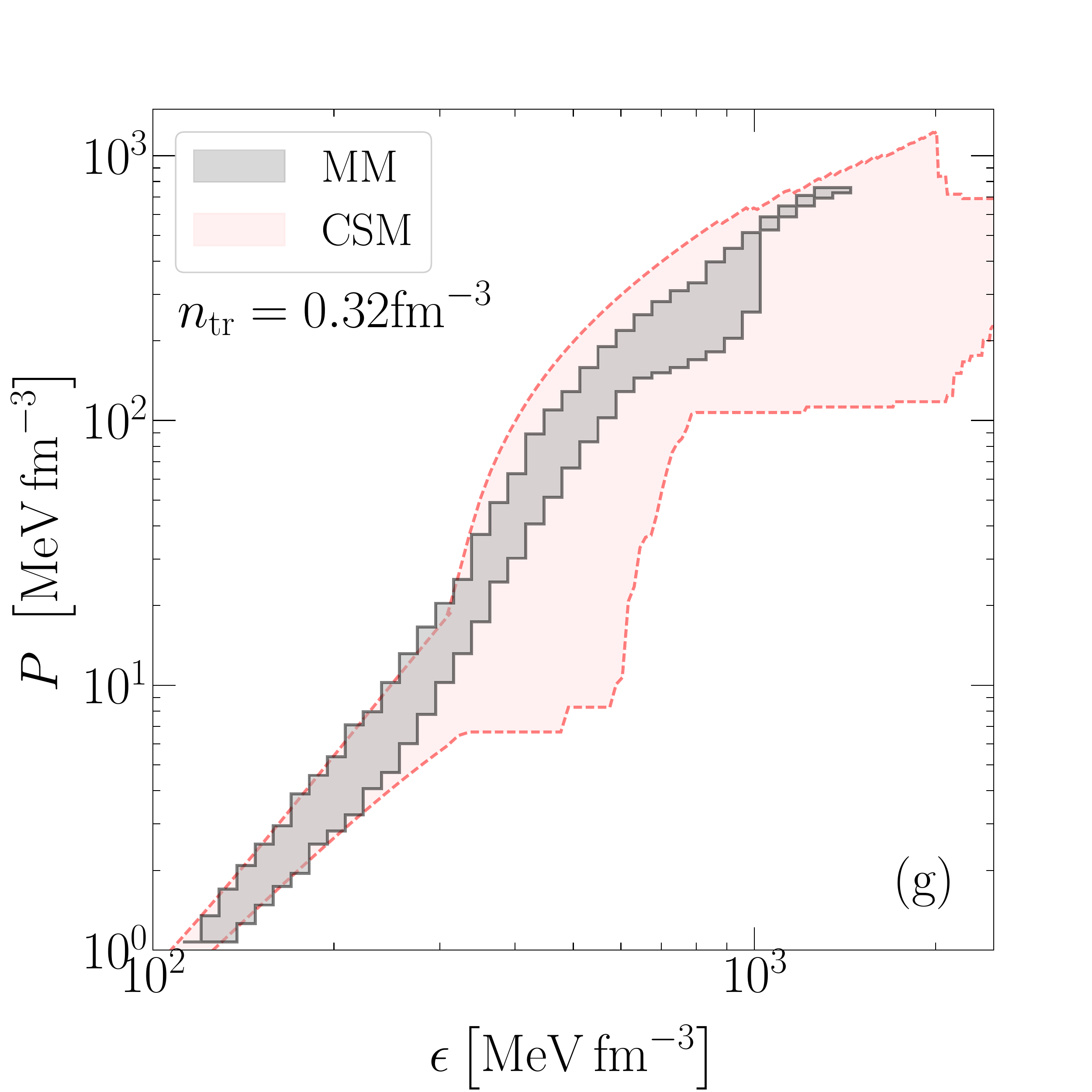}
\includegraphics[trim= 0.0cm 0 0 0, clip=,width=0.65\columnwidth]{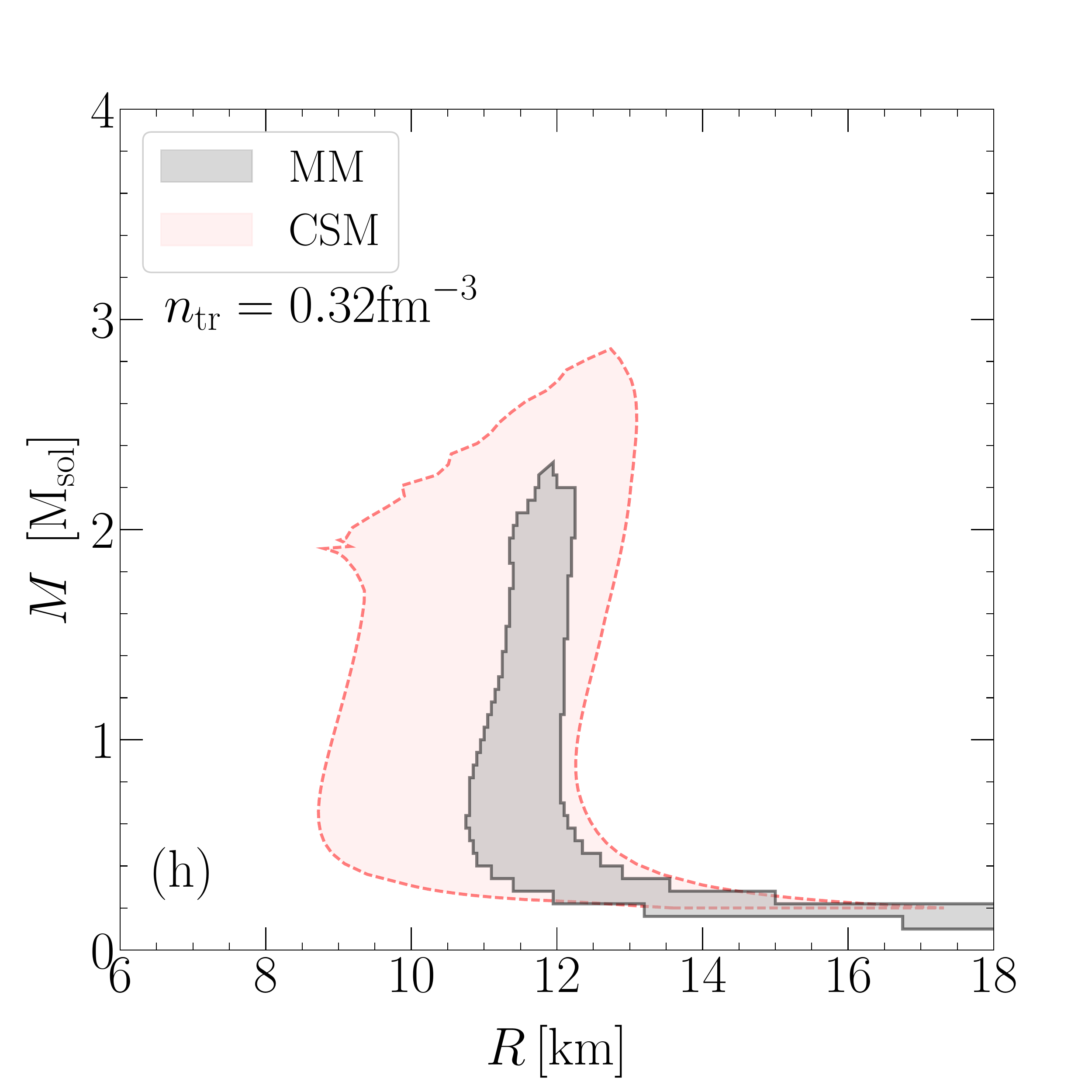}
\caption{\label{fig:EOSMRcomp}
(a)-(d) Comparison of the MM (black) and the CSM (red) for $n_{\text{tr}}=n_{\text{sat}}$ and (e)-(h) $n_{\text{tr}}=2 n_{\text{sat}}$. We compare (a), (e) the energy per particle and (b), (f) pressure in PNM, (c), (g) the EOS envelopes of NS matter,  and (d), (h) the MR envelopes. In panels (a), (b), (e), and (f), we also show the PNM constraints from Ref.~\cite{Tews:2018kmu}.
}
\end{figure*}

The energy per particle and pressure of PNM, determined from chiral EFT~\cite{Tews:2018kmu}, are shown as functions of the baryon number density $n$ in Figs.~\ref{fig:EOSMRcomp}(a) and \ref{fig:EOSMRcomp}(b) up to $n_\text{sat}$ and in Figs.~\ref{fig:EOSMRcomp}(e) and \ref{fig:EOSMRcomp}(f) up to $2n_\text{sat}$ (vertical blue uncertainty bars). 

\subsection{Minimal model}

As mentioned earlier, for $n>n_\text{tr}$ we use two distinct approaches. In the first approach, which we call the minimal model, the EOS is assumed to be smooth and related to a density expansion about $n_{\text{sat}}$. The model parameters, which can in principle be measured in experiments, are the so-called empirical parameters of nuclear matter: the saturation energy $E_{\text{sat}}$, the incompressibility $K_{\text{sat}}$, the symmetry-energy parameter $E_{sym}$, its slope parameter $L_{sym}$, and higher-order parameters, defined through 
\begin{align}
e_s(n) &= E_{\text{sat}} + \frac 1 2 K_{\text{sat}} u^2 + \frac 1 6 Q_{\text{sat}} u^3 \nonumber  \label{eq:esat}\\ 
& + \frac 1 {24} Z_{\text{sat}} u^4 + ... \\
s(n) &= E_{\text{sym}} + L_{\text{sym}} u+ \frac{1}{2} K_{\text{sym}} u^2 + \frac{1}{6} Q_{\text{sym}} u^3 \nonumber  \label{eq:esym}\\ 
& +\frac{1}{24} Z_{\text{sym}} u^4 + ... 
\end{align}
with the energy per particle in symmetric matter $e_s(n)$ and the symmetry energy $s(n)=\frac 1 2\partial^2 e/ \partial \delta^2$. Furthermore, the expansion parameter is defined as $u=(n-n_{\text{sat}})/(3n_{\text{sat}})$ with the baryon density $n=n_n+n_p$ and the neutron and proton densities $n_{n}$ and $n_{p}$, respectively. Lastly, the isospin asymmetry is given by $\delta=(n_n-n_p)/n$. By varying the empirical parameters within their uncertainties, the MM is able to reproduce the EOSs predicted by a large number of existing models which assume a nuclear description for all densities encountered in NSs~\cite{Margueron:2017eqc}. Specifically, the MM corresponds to the meta-model ELFc introduced and applied to NS in Refs.~\cite{Margueron:2017eqc, Margueron:2017lup}.

\subsection{Speed-of-sound model}

We use a second approach to extend the EOS to higher densities in terms of the speed of sound, and we call this model CSM. The CSM constructs the EOS from a general parametrization of the speed of sound, $c_S^2 = \partial P(\epsilon)/\partial \epsilon$, with pressure $P$ and energy density $\epsilon$. It includes phase transitions that can produce drastic softening or stiffening, contrasting the smoothness imposed by the MM. The CSM explores the widest possible domain of equations of state, maximizing the predictions for the EOS and neutron-star observables. It was studied in Ref.~\cite{Tews:2018kmu}, but here we extend the parametrization to explore the full space for $c_S$ by randomly sampling six reference points $c_S(n)$ between $n_{\rm tr}$ and $12 n_{\text{sat}}$ and connecting them by linear segments. For each such EOS, we generate an additional EOS by including a strong first-order phase transition with random position and width. We have checked the stability of the resulting envelopes of this extension shown in Fig.~\ref{fig:EOSMRcomp} against increasing the number of reference points to ten. 

This approach represents a generalization of extensions that use piecewise polytropes, introduced in Ref.~\cite{Read:2008iy} and widely used for NSs~\cite{Hebeler:2010,Hebeler:2013nza, Raithel:2016bux, Annala:2017llu}, but offers the advantage that $c_S$, entering in the calculation of $\Lambda$, is continuous except when first-order phase transitions are explicitly considered. Similar to piecewise polytropes, the CSM does not allow to extract information on the composition of dense matter or the type of a phase transition but is a pragmatic way to test effects beyond the MM, and allows us to discuss the observable differences between nucleonic and exotic phases of matter. 

\subsection{Neutron-star equation of state}

To obtain the neutron-star equations of state, we extend our models to $\beta$ equilibrium and include a crust, as described in Refs.~\cite{Margueron:2017lup, Tews:2016ofv}. These different prescriptions show excellent agreement, see Figs.~\ref{fig:EOSMRcomp}(c) and \ref{fig:EOSMRcomp}(g) at low densities. For all our EOSs, once $\beta$-equilibrium matter is obtained, we enforce causality, stability, and, in the MM, the positivity of the symmetry energy, $s(n)>0$, up to the maximal central density corresponding to $M_{\text{max}}$. Current pulsar observations impose the additional conservative constraint $M_{\text{max}}>1.9M_\odot$~\cite{Demorest2010,Antoniadis:2013pzd,Fonseca2016}.

In Fig.~\ref{fig:EOSMRcomp}(c), we compare the EOS boundaries for the CSM (light-red bands, dashed contours) and the MM (dark-gray bands, solid contours) for $n_{\text{tr}}=n_{\text{sat}}$ and in Fig.~\ref{fig:EOSMRcomp}(g) for $n_{\text{tr}}=2 n_{\text{sat}}$. In Figs.~\ref{fig:EOSMRcomp}(d) and \ref{fig:EOSMRcomp}(h) we show the corresponding boundaries in the mass-radius (MR) diagram, resulting from solving the Tolman-Oppenheimer-Volkoff equations~\cite{Tolman,OppenheimerVolkoff}. We find that (i) both models show good agreement at low densities, (ii) the MM is a subset of the CSM above $n_\text{tr}$ as expected, and (iii) the CSM allows for regions of sudden stiffening or softening that are absent in the MM. From Fig.~\ref{fig:EOSMRcomp}, we also conclude that (iv) although the PNM EOS has sizable uncertainties between $1-2\, n_{\text{sat}}$, see Ref.~\cite{Tews:2018kmu} and Figs.~\ref{fig:EOSMRcomp}(e) and \ref{fig:EOSMRcomp}(f), it nevertheless provides sufficient additional information to substantially reduce the EOS and MR uncertainties; see also Ref.~\cite{Lattimer:2000nx}. 

\section{Implications of GW170817}\label{sec:GW170817}
\subsection{Tidal polarizabilities}

\begin{figure}[t]
\includegraphics[trim= 0.2cm 0.1cm 0.3cm 0.3cm, clip=,width=0.9\columnwidth]{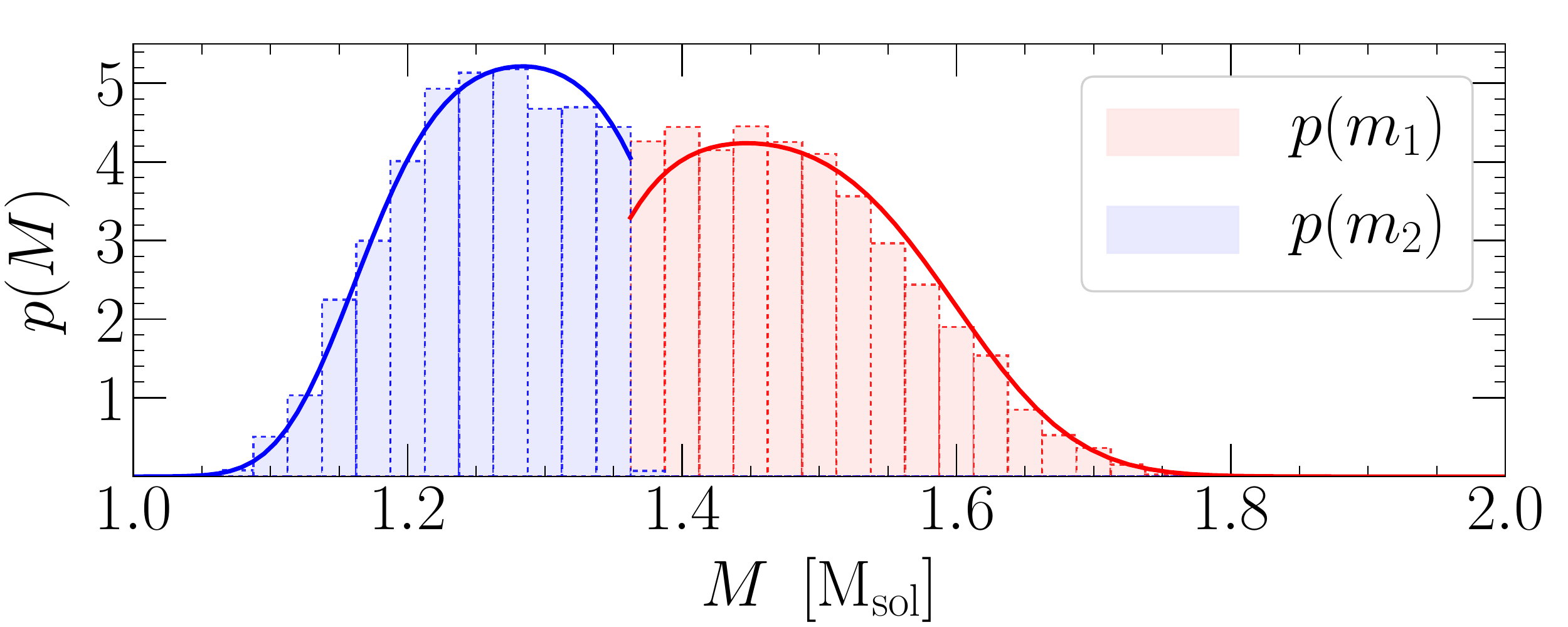}
\caption{\label{fig:M1M2histo}
The mass distributions for $m_1$ and $m_2$ from Ref.~\cite{Abbott:2018wiz} (histograms) and the distributions used in this work (solid lines).
}
\end{figure}

We now discuss the implications of GW170817. The wavefront analysis of the entire signal provided very tight constraints on the chirp mass of GW170817, defined as $M_{\text{chirp}}=(m_1 m_2)^{3/5}m_{\text{tot}}^{-1/5}$ with $m_{\text{tot}}=m_1+m_2$, and less tight constraints on the mass distribution of the two neutron stars. With $m_1\, (m_2)$ being the mass of the heavier (lighter) neutron star, the mass asymmetry can be parametrized by the ratio $q=m_2/m_1$. 
The observational constraints on $M_{\text{chirp}}$ and $q$ can be described by an analytical probability distribution~\cite{Margalit:2017} ,
\begin{equation}
p(q,M_{\text{chirp}}) = p(q) p(M_{\text{chirp}})\,,
\end{equation}
where
\begin{equation}
p(M_{\text{chirp}}) \propto \exp [- (M_{\text{chirp}}-\bar{M}_{\text{chirp}})^2/2\sigma_{M}^2]\,,
\end{equation}
with $\bar{M}_{\text{chirp}}=1.186M_{\odot},\,\sigma_{M}=10^{-3}M_{\odot}$~\cite{Abbott:2018wiz},
and
\begin{equation}
p(q)=\exp \left(-\frac12 v(q)^2 -\frac{1.83}{2} v(q)^4 \right)\,,
\end{equation}
with $v(q)=(q-0.89)/0.20$.
We compare the normalized observed and analytical mass distributions for $m_1$ and $m_2$, sampled from $p(q,M_{\text{chirp}})$, in Fig.~\ref{fig:M1M2histo}. Note that here we only investigate the more realistic low-spin case, because large spins are not expected from the observed galactic binary NS population; see also Ref.~\cite{Abbott:2018exr}.

The combined tidal parameter
\begin{equation}
\tilde{\Lambda}~=~\frac{16}{13} \left[\frac{(m_1+12m_2)m_1^4\Lambda_1 }{m_{\text{tot}}^5}+ 1 \leftrightarrow 2\right]
\end{equation}
appears as a post-Newtonian fifth-order correction to the wavefront phase~\cite{Flanagan:2007ix,Damour:2009vw}, which makes it small and therefore difficult to extract from GW observations~\cite{Favata:2013rwa,Yagi:2013baa}.
For GW170817, limits of $70\le \tilde{\Lambda}\le 720$ were reported~\cite{Abbott:2018wiz} (hereafter LV constraint). Given the sizable observational uncertainty for $\tilde{\Lambda}$, we now analyze $\tilde{\Lambda}$ for GW170817 as predicted by nuclear physics. Using the two general EOS models and the mass distributions as specified before, we compute the correlation for $\Lambda_1$ and $\Lambda_2$. We present the envelopes for the CSM (top half) and the MM (lower half, $\Lambda_1\leftrightarrow\Lambda_2$) in Fig.~\ref{fig:LamLam}, and compare their respective 50\% and 90\% probability contours with the corresponding LV results. 

In Fig.~\ref{fig:LamLam}(a), we show results for $n_{\text{tr}}=n_{\text{sat}}$. Due to the large uncertainty in the high-density EOS, the contour for the CSM is much wider than the LV contour, and we find 
$\tilde{\Lambda}_{\text{CSM}}=60-2180$, compared with 
$\tilde{\Lambda}_{\text{MM}}=280-1030$. In this case, the LV constraints can provide additional information for the EOS of NSs (see also Refs.~\cite{Annala:2017llu, Most:2018hfd}).
 
When we enforce the LV constraint by selecting appropriate EOS-$q$ combinations, we obtain the envelopes presented in Fig.~\ref{fig:LamLam}(b). In this case, our two models are in very good agreement with each other. This agreement confirms previous works~\cite{Abbott:2017,Annala:2017llu, Paschalidis:2017qmb, Most:2018hfd}, showing that the LV constraint excludes the stiffest EOSs leading to less compact NSs. In our case, an $M=1.4M_\odot$ NS has a radius $9.0<R_{1.4}<13.6$~km ($11.3 <R_{1.4}< 13.6$~km for the MM), similar to Ref.~\cite{Annala:2017llu}.

Finally, in Fig.~\ref{fig:LamLam}(c), we show our results for $n_{\text{tr}}=2n_{\text{sat}}$, and no $\tilde{\Lambda}$ constraint. In this case, the envelopes are much narrower than the LV result. This highlights the fact that, even though the neutron-matter EOS has sizable uncertainties at $2n_\text{sat}$, nuclear-physics calculations provide sufficient information to decrease uncertainties for $\tilde{\Lambda}$ below current observational limits. In this case, we predict 
$\tilde{\Lambda}_{\text{CSM}}=80-580$ and 
$\tilde{\Lambda}_{\text{MM}}= 280-480$, and a 
$M=1.4M_\odot$ NS has a radius $ 9.2 <R_{1.4}< 12.5$~km ($11.3 <R_{1.4}< 12.1$~km for the MM).

While the CSM can also produce small values for $\tilde{\Lambda}$ in all cases shown in Fig.~\ref{fig:LamLam}, the MM prevents $\tilde{\Lambda}\lesssim 250$. This quantitative difference is directly related to the possibility of phase transitions in the CSM, where significant softening of the EOS in the vicinity of $2n_\text{sat}$ followed by a stiffer EOS at higher density produces compact neutron stars while still accounting for the observed $M_{\text{max}}$.  

\begin{figure}[t]
\includegraphics[trim= 0.0cm 0 0 0, clip=,width=0.9\columnwidth]{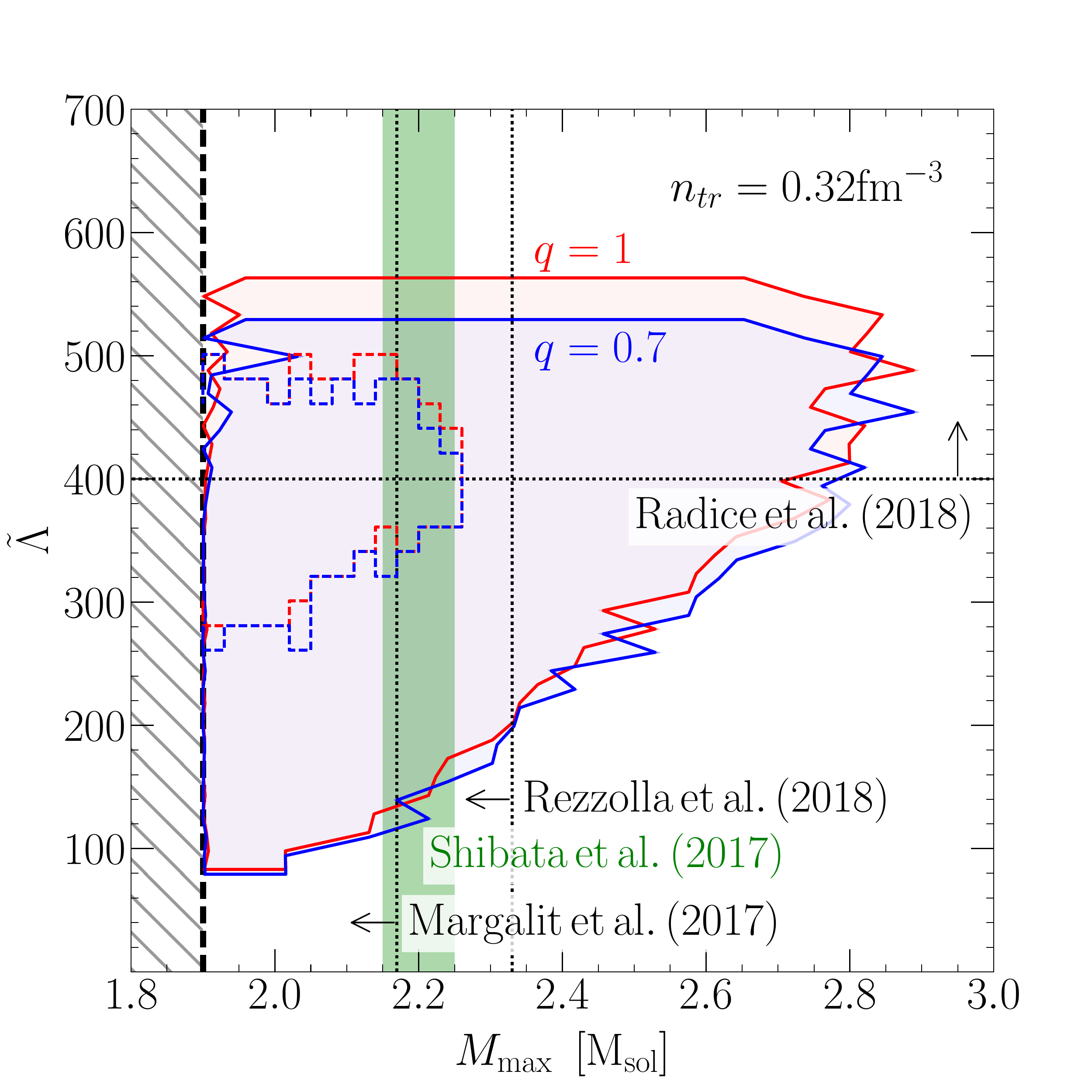}
\caption{\label{fig:MmaxLam}
Envelopes for the $\tilde{\Lambda}-M_{\text{max}}$ correlation for $q=1$ (red) and $q=0.7$ (blue) for $n_{\text{tr}}=2 n_{\text{sat}}$ for the CSM (solid) and the MM (dashed). We also show the $M_{\text{max}}$ constraints of Refs.~\cite{Shibata:2017xdx,Margalit:2017,Rezzolla:2017aly} and the $\tilde{\Lambda}$ constraint of Ref.~\cite{Radice:2017lry}.
}
\end{figure} 

\subsection{$\tilde{\Lambda}-M_{\text{max}}$ correlation}

Finally, we study the $\tilde{\Lambda}-M_{\text{max}}$ correlation for $n_\text{tr}=2n_\text{sat}$. In Fig.~\ref{fig:MmaxLam} we present envelopes of our predictions for $q=1$ and $q=0.7$, corresponding to the upper and lower limits for GW170817~\citep{Abbott:2018wiz}. The MM domain in $\tilde{\Lambda}-M_{\text{max}}$ is much smaller than for the CSM, as anticipated from Fig.~\ref{fig:LamLam}. For the MM, $M_{\text{max}}$ is very compatible with the inferences of about $2.2M_\odot-2.3M_\odot$~\cite{Shibata:2017xdx,Margalit:2017,Rezzolla:2017aly}, but for either model an upper bound for $M_{\text{max}}$ does not constrain $\tilde{\Lambda}$.

In Ref.~\cite{Radice:2017lry}, merger simulations using four different equations of state and several values for $q$ were performed and the correlation between  $\tilde{\Lambda}-M_{\text{max}}$ was used to argue that $\tilde{\Lambda}>400$ under the assumption that the hypermassive neutron star in GW170817 did not collapse promptly to a black-hole. From Fig.~\ref{fig:MmaxLam}, it is clear that there exist equations of state with $\tilde{\Lambda}<400$ but much higher $M_{\text{max}}$ than the four EOSs used in Ref.~\cite{Radice:2017lry}.  For example, there exist EOSs with $\tilde{\Lambda}=300$ but with $M_{\text{max}}=2.6 M_{\odot}$, large enough to support a stable long-lived merger remnant. We therefore suggest that the lower limit of Ref.~\cite{Radice:2017lry} might be overestimated.
 
\section{Summary}\label{sec:Summary}
 
In conclusion, we have shown that neutron-matter calculations using chiral EFT interactions up to $2 n_{\text{sat}}$ predict
$\tilde{\Lambda}_{\text{CSM}}=80-580$ and 
$\tilde{\Lambda}_{\text{MM}}=280-480$ -- a range that is consistent but smaller than the constraints obtained from the
LV analysis of data from GW170817 \cite{Abbott:2018wiz}.
This suggests that future GW detections will need to be more precise to provide useful EOS constraints. 

Furthermore, we have contrasted two high-density extrapolations (MM and CSM) to provide insights on how measurements of $\tilde{\Lambda}$ from binary neutron-star mergers can elucidate properties of matter at supra-nuclear densities. For example, a detection of $\tilde{\Lambda}$ outside the range of 
$280-480$ would provide strong evidence for the existence of phase transitions at supra-nuclear densities. Our analysis also suggests that, if future detections favor values incompatible with 
$\tilde{\Lambda}_{\text{CSM}}=80-580$, it would imply a 
breakdown of the nuclear EFT between $1-2 n_{\text{sat}}$, and perhaps signal the presence of strongly-interacting dense quark matter inside neutron stars.  

Ongoing efforts to improve the EFT predictions for neutron matter, especially those pertaining to three-neutron forces, will likely reduce the present range for $\tilde{\Lambda}_{\text{CSM}}$ and $\tilde{\Lambda}_{\text{MM}}$, and improve prospects for GW detections to provide unique insights into the 
nature of dense matter in NS. 

\begin{acknowledgments}
We thank the participants of the INT-JINA Symposium "First multimessenger observations of a neutron-star merger and its implications for nuclear physics" for useful discussions. 
This work was supported in part by
the National Science Foundation Grant No.~PHY-1430152 
(JINA Center for the Evolution of the Elements) and
the U.S.~DOE under Grants No.~DE-FG02-00ER41132.
JM was partially supported by the IN2P3 Master Project MAC, "NewCompStar" COST Action MP1304, and PHAROS COST Action MP16214.
\end{acknowledgments}

\bibliography{final}{}

\end{document}